\documentclass[twocolumn,aps,prd]{revtex4-2}
 \usepackage{graphicx,amssymb}
 \usepackage[utf8]{inputenc} 
 
\usepackage[urlcolor=blue,colorlinks,breaklinks=true]{hyperref}

\PassOptionsToPackage{obeyspaces,spaces,hyphens}{url}

 \usepackage{amsmath}

\begin{document}
 	\def\half{{1\over2}}
 	\def\shalf{\textstyle{{1\over2}}}
 	
 	\newcommand\lsim{\mathrel{\rlap{\lower4pt\hbox{\hskip1pt$\sim$}}
 			\raise1pt\hbox{$<$}}}
 	\newcommand\gsim{\mathrel{\rlap{\lower4pt\hbox{\hskip1pt$\sim$}}
 			\raise1pt\hbox{$>$}}}

\newcommand{\be}{\begin{equation}}
\newcommand{\ee}{\end{equation}}
\newcommand{\bq}{\begin{eqnarray}}
\newcommand{\eq}{\end{eqnarray}}
 	
\title{Analytical scaling solutions for the evolution of cosmic domain walls in a parameter-free velocity-dependent one-scale model}
 	 	
\author{P.P. Avelino}
\email[Electronic address: ]{pedro.avelino@astro.up.pt}
\affiliation{Departamento de F\'{\i}sica e Astronomia, Faculdade de Ci\^encias, Universidade do Porto, Rua do Campo Alegre 687, PT4169-007 Porto, Portugal}
\affiliation{Instituto de Astrof\'{\i}sica e Ci\^encias do Espa{\c c}o, Universidade do Porto, CAUP, Rua das Estrelas, PT4150-762 Porto, Portugal}

\author{D. Gr{\"u}ber}
\email[Electronic address: ]{david.grueber@astro.up.pt}
\affiliation{Departamento de F\'{\i}sica e Astronomia, Faculdade de Ci\^encias, Universidade do Porto, Rua do Campo Alegre 687, PT4169-007 Porto, Portugal}
\affiliation{Instituto de Astrof\'{\i}sica e Ci\^encias do Espa{\c c}o, Universidade do Porto, CAUP, Rua das Estrelas, PT4150-762 Porto, Portugal}

\author{L. Sousa}
\email[Electronic address: ]{lara.sousa@astro.up.pt}
\affiliation{Departamento de F\'{\i}sica e Astronomia, Faculdade de Ci\^encias, Universidade do Porto, Rua do Campo Alegre 687, PT4169-007 Porto, Portugal}
\affiliation{Instituto de Astrof\'{\i}sica e Ci\^encias do Espa{\c c}o, Universidade do Porto, CAUP, Rua das Estrelas, PT4150-762 Porto, Portugal}
\date{\today}

\date{\today}
\begin{abstract}
We derive an analytical approximation for the linear scaling evolution of the characteristic length $L$ and the root-mean-squared velocity $\sigma_v$ of standard frictionless domain wall networks in Friedmann-Lemaître-Robertson-Walker universes with a power law evolution of the scale factor $a$ with the cosmic time $t$ ($a \propto t^\lambda$). This approximation, obtained using a recently proposed parameter-free velocity-dependent one-scale model for domain walls,  reproduces well the model predictions for $\lambda$ close to unity,  becoming exact in the $\lambda \to 1^-$ limit. We use this approximation, in combination with the exact results found for $\lambda=0$, to obtain a fit to the model predictions valid for $\lambda \in [0, 1[$ with a maximum error of the order of $1 \%$. This fit is also in good agreement with the results of field theory numerical simulations, specially for $\lambda \in [0.9, 1[$. Finally, we explicitly show that the phenomenological energy-loss parameter of the original velocity-dependent one-scale model for domain walls vanishes in the $\lambda \to 1^-$ limit and discuss the implications of this result.
\end{abstract}

\maketitle
 	
\section{Introduction}
\label{sec:intr}

The production of cosmic defects through the Kibble mechanism is a generic prediction of grand unified scenarios \cite{Kibble:1976sj}. These defect networks can leave behind a variety of observable cosmological signatures, provided that they survive long enough and their characteristic energy scale is sufficiently high (see, for example, \cite{2000csot.book.....V} and references therein). An accurate description of the cosmological evolution of cosmic defects is crucial to perform accurate predictions of their observational consequences. This is usually done either by resorting to Nambu-Goto~\cite{Bennett:1987vf,Albrecht:1989mk,Allen:1990tv,Blanco-Pillado:2011egf} or field theory numerical simulations~\cite{Vincent:1997cx,1989ApJ...347..590P,Hindmarsh:2014rka,Martins:2016ois}, or by using semianalytical models describing the evolution of a few thermodynamic variables which characterize the large-scale dynamics of the networks~\cite{Martins:1996jp,Martins:2000cs,Sousa:2011ew,Sousa:2011iu}. While the first approach might be preferable for specific models inspired by particle physics, the latter is more versatile since it can easily accommodate a variety of well motivated scenarios or even the phenomenological modeling of a priori unknown physics. These semianalytical models have been used with great success to constrain cosmological scenarios with cosmic strings~\cite{Pogosian:1999np, Wyman:2005tu,Charnock:2016nzm, Sousa:2016ggw, Guedes:2018afo, Auclair:2019wcv, Sousa:2020sxs, Rybak:2021scp} and domain walls~\cite{Avelino:2014xsa,Sousa:2015cqa}.

A velocity-dependent one-scale (VOS) model for the evolution of the characteristic length $L$ and of the root-mean-squared velocity $\sigma_v$ of domain wall networks in flat expanding or collapsing homogeneous and isotropic universes was developed in~\cite{Avelino:2010qf,Avelino:2011ev}, generalizing previous work on cosmic strings~\cite{Martins:1996jp,Martins:2000cs} (see also for~ \cite{Sousa:2011ew,Sousa:2011iu,Avelino:2015kdn} for a unified description of $p$-brane dynamics). This model has been shown to provide an accurate description of domain wall dynamics not only in cosmology, but also in the context of non-relativistic systems in condensed matter \cite{Avelino:2010qf} and biology \cite{2012PhRvE..86c1119A} wherein the dynamics is dominated by the curvature of domain interfaces. Unfortunately, this approach relies on numerical simulations for the (cosmology-dependent) calibration of its two phenomenological parameters, usually referred to as energy-loss and momentum parameters (see also~\cite{Martins:2016ois,Rybak2018} for a six parameter extension of the standard parametric domain wall VOS model).

Recently, the evolution of cosmological domain walls in an expanding universe characterized by a power law evolution of the scale factor $a$ with the cosmic time $t$ ($a \propto t^\lambda$) has been investigated using a parameter-free VOS model \cite{Avelino:2019wqd,Avelino:2020ubr}. The model predictions have been compared with the results of field theory numerical simulations, with a notable agreement obtained for $\lambda \in [0.9, 1[$. Although for slower expansion rates the values of the characteristic length $L$ and root-mean-squared velocity $\sigma_v$ predicted by the parameter-free VOS model can be larger by up to $30 \, \%$ than the reported numerical results, a number of potential problems with the determination of $L$ and $\sigma_v$ in field theory numerical simulations have been identified. These problems are expected to manifest themselves especially in the relativistic regime~\cite{Hindmarsh:2014rka,Avelino:2019wqd}, and need to be addressed before a more meaningful comparison with the model predictions can be performed.

In the present paper we use the parameter-free VOS model to derive an analytical approximation for the evolution of the characteristic length $L$ and root-mean-squared velocity $\sigma_v$ of standard frictionless domain wall networks in homogeneous and isotropic  Friedmann-Lemaître-Robertson-Walker (FLRW) universes undergoing a power law evolution of the scale factor $a$ with cosmic time $t$ ($a \propto t^\lambda$, with constant $\lambda$).  In Sec. \ref{sec2} we briefly review the parameter-free VOS model for the evolution of cosmological domain walls. In Sec. \ref{sec3} we derive several analytical results for the dynamics of cylindrical and spherical domain walls and use them to determine the linear scaling parameters of the parameter-free VOS model for $\lambda=0$ and $\lambda \to 1^-$. We also briefly discuss the curvature-dominated evolution of individual domains in the $\lambda \to 1^-$ limit. In Sec. \ref{sec4} we shall use the approximation found for $\lambda$ close to unity, in combination with the exact results found for $\lambda=0$, to obtain a fit to the model predictions valid for $\lambda \in [0, 1[$ with a maximum error of the order of $1 \%$. We also confront the predictions of the model with the results of numerical simulations.
In Sec. \ref{sec5} we provide an alternative demonstration of the dependence of the linear scaling parameters on $\lambda$, in the $\lambda \to 1^-$ limit, using the standard VOS model for domain walls and determine the value of the phenomenological energy-loss parameter in this limit. We also discuss the implications of this result for more general multi-parameter domain wall VOS models.  Finally, we conclude in \ref{sec6}.

We shall use fundamental units with $c=1$, where $c$ is the speed of light in vacuum.

\section{Parameter-free domain wall VOS model}
\label{sec2}

In this section we briefly describe the parameter-free domain wall VOS model for the evolution of standard frictionless domain wall networks proposed in~\cite{Avelino:2019wqd}. We shall consider a flat $3+1$-dimensional homogeneous and isotropic FLRW universe with line element
\be
d s^2 =a^2[\eta] \left(d \eta^2 - d {\bf q} \cdot d {\bf q}  \right) \,,
\ee
where $a$ is the cosmological scale factor, $\eta=\int dt/a$ is the conformal time, $t$ is the physical time and $\bf q$ are comoving spatial coordinates. In this paper we focus on expanding cosmologies having a power law evolution of the scale factor with the physical time: $a \propto t^\lambda$ with $\lambda \in \ [0,1[$, so that 
\be
a \propto \eta^{\lambda/(1-\lambda)}\,.\label{lambda}
\ee.

Our parameter-free VOS model~\cite{Avelino:2019wqd} is motivated by field theory simulations of standard domain wall network evolution which show that domain wall intersections are rare~\cite{Martins:2016ois} and by the fact that thin domain walls are not expected to produce significant amounts of scalar radiation, except in the final stages of collapse~\cite{Vachaspati:1984yi}. In this model, the network is assumed to be composed of infinitely thin domain walls that are taken to be either all spherical ($s=2$) or all cylindrical ($s=1$) --- for $s=1$ all the domain walls are assumed to be oriented along parallel axes. The corresponding energy or energy per unit length (for spherical or cylindrical walls respectively) is equal to 
\be
E [\eta,q]= 2 \pi s \sigma_{\rm w0} a^s q^s \gamma\,,
\ee
where, $q$ represents the comoving radius of each domain wall, $\gamma=(1-v^2)^{-1/2}$, $v=|dq/d\eta|$, and $\sigma_{\rm w0}$ is the proper domain wall energy per unit area. The consideration of both spherical and cylindrical configurations makes it possible to evaluate the impact of geometry on the model predictions. 

Domain walls are assumed to start at rest at some early conformal time $\eta_{\rm i}$ with an initial comoving size $q_{\rm i}$ (which varies from wall to wall) and therefore to never intersect each other. When a domain wall reaches $q=0$ it is assumed to decay instantaneously, ceasing to be part of the network. The initial comoving radius of the domain walls that disappear at a conformal time $\eta$ shall be denoted by $q_{\rm i*}$. The probability density function of the initial comoving radii $q_{\rm i}$ of the domain walls is given by \cite{Avelino:2019wqd}
\be
{\mathcal P} \equiv {\mathcal P} [q_{\rm i}] =(1+s) \eta_{\rm i}^{1+s} q_{\rm i}^{-2-s} \Theta[q_{\rm i}-\eta_{\rm i}] \,,
\ee
where $\Theta$ is the Heaviside step function. The form of this probability density function was chosen in such a way that the initial energy density of domain walls with $q_{\rm i}$ larger than $\eta > \eta_{\rm i}$ satisfies 
\be
\rho_{\rm w i} (q_{\rm i} > \eta) = \int_{\eta}^\infty    {\mathcal P}[q_{\rm i}]  q_{\rm i}^s dq_{\rm i} \propto \eta^{-1} \,.
\ee
It was also required that it satisfies the normalization condition  $\int_0^\infty {\mathcal P} [q_{\rm i}] dq_{\rm i} =1$.

The total energy density of domain walls at a conformal time $\eta \gg \eta_{\rm i}$ can be computed as
\be
\rho_{\rm w}  = \int_{q_{\rm i*}}^\infty n[\eta] E[\eta,q_{\rm i}]  {\mathcal P}[q_{\rm i}] dq_{\rm i} \,.
\label{rhodwi}
\ee
Here, $q_{\rm i*}$ is the minimum initial comoving radius required for a domain wall to survive until the conformal time $\eta$, $n= n_{\rm i} a^{-1-s}$, where $n_{\rm i}$ is the initial domain wall number density defined as the number of walls per unit volume (spherical case) or per unit area (cylindrical case), and the scale factor is normalized to unity at the initial time ($a_{\rm i} \equiv a(\eta_{\rm i}) =1$). 

The world history of an infinitely thin featureless domain wall in a flat expanding FRLW universe can be represented by a three-dimensional world-sheet obeying the usual Nambu-Goto action. For a cylindrical or a spherical domain wall ($s=1$ or $s=2$, respectively) and a power law expansion described by Eq. \eqref{lambda}, the corresponding equations of motion may be written as
\be
\frac{d\tilde q}{d \tau}=-v \,, \qquad \frac{d v}{d\tau}+\gamma^{-2}\left( \frac{3 \lambda}{1-\lambda}\frac{v}{\tau} - \frac{s}{\tilde q}\right)=0\,, 
\label{dwev}
\ee
where $\tau=\eta/q_{\rm i}$, $\tilde q [\tau]=q/q_{\rm i}$, and $v=v[\tau]$.

The characteristic length of the network, defined by $L \equiv \sigma_{\rm w0}/\rho_w$, satisfies \cite{Avelino:2019wqd}
\be
\zeta \equiv \frac{L}{a\eta}=\left(\beta \int_0^{\tau_*} {\widetilde q}^s \gamma d \tau\right)^{-1}\,, \label{zetadef}
\ee
where $\beta=2 \pi s (1+s)  n_{\rm i} \eta_{\rm i}^{1+s}$ and $\tau_*=\eta/q_{\rm i*}$, while the mean-squared velocity of the domain walls is given by \cite{Avelino:2019wqd}
\be
\sigma_v^2 \equiv  \frac{\int_{q_{\rm i*}}^\infty v^2 n E {\mathcal P} dq_{\rm i}}{ \int_{q_{\rm i*}}^\infty n E  {\mathcal P} dq_{\rm i} }  = \frac{\int_0^{\tau_*} v^2 {\widetilde q}^s \gamma d \tau}{\int_0^{\tau_*} {\widetilde q}^s \gamma d \tau}\,. \label{svdef}
\ee

\section{Analytical results for $\zeta$ and $\sigma_v$}
\label{sec3}

Here, we shall derive analytical approximations for $\zeta$ and $\sigma_v$ in two limiting cases: $\lambda \to 1^-$ and $\lambda=0$. For $\lambda$ sufficiently close to unity, the domain walls are almost always non-relativistic, making it possible to perform an analytical estimation of both parameters. On the other hand, the absence of Hubble damping in the $\lambda=0$ case also allows for an analytical determination of both $\zeta$ and $\sigma_v$.

\subsection{The $\lambda \to 1^-$ case}
\label{non-rel-app}

Consider the evolution of the domain walls in the non-relativistic regime.  An approximate solution of Eq.~(\ref{dwev}), valid for $v \ll 1$, 
is given by
\bq
\gamma^{\frac{1}{s}} v & = &  \frac{1}{\tau_*^2}\frac{\tau}{\tilde q}\,, \label{analytic1a}\\
\tilde q  & = & \sqrt{1-\left( \frac{\tau}{\tau_*}\right)^2} \,, \label{analytic1}
\eq
with
\be
\tau_*=\sqrt{\frac{2\lambda+1}{s(1-\lambda)}}\,.
\label{analytic2}
\ee
Although the factor of $\gamma^{1/s}$ in Eq. (\ref{analytic1a}) is very close to unity in the non-relativistic regime, we have included it in order to ensure that $E \propto {\tilde q}^s \gamma$ tends to a non-vanishing constant in the $\tau \to \tau_*$ limit. Since $v$ can never be larger than unity, this condition ensures that the energy of the domain wall is always well defined, thus allowing for the use of the approximation outside the strict non-relativistic limit assumed in its derivation. Also notice that $\tau_*$, the value of $\tau$ for which $\tilde q=0$, is strongly dependent on $\lambda$: $\tau_* \propto (1-\lambda)^{-1/2}$ for $\lambda \sim 1$, becoming infinite in the $\lambda \to 1^-$ limit.  It also shows some dependence on whether the domain walls are  assumed to be cylindrical or spherical, being smaller by a factor of $\sqrt 2$ in the latter case. 

Let us define $\tau_+$ as the value of $\tau$ for which $v \gamma^{1/s}=1$. One can show that, according to our approximation, 
\be
\frac{\Delta \tau}{\tau_*}=\frac{\tau_*-\tau_+}{\tau_*}\sim \frac{1}{1+\tau_*^2}\,.
\ee
This essentially means that, for $\lambda$ sufficiently close to unity, the fraction of the conformal lifetime of the domain walls in which they are relativistic can be made arbitrarily close to zero ($\Delta \tau \sim \tau_*^{-1} \propto \sqrt{1-\lambda}$ for $\lambda$ close to unity). 
This enables us to roughly estimate the contribution to the integrals given in Eqs. (\ref{zetadef}) and (\ref{svdef}) coming from the relativistic regime. In the late stages of collapse $v \sim 1$, which implies, according to Eq. (\ref{analytic1a}), that both $\tilde{q}^s\gamma$ and  $v^2\tilde{q}^s\gamma$ are approximately equal to $\tau^s \tau_*^{-2s} \sim \tau_*^{-s}$ in this regime for $\lambda$ sufficiently close to unity. Evaluating the aforementioned integrals in the last stages of collapse (i.e., between $\tau_+$ and $\tau_*$), we find that the relativistic contribution is of order $\Delta \tau \, \tau_*^{-s}  \sim   \tau_*^{-1-s}  \propto (1-\lambda)^{(1+s)/2}$ and is thus expected to become negligible in the $\lambda \to 1^-$ limit. Hence, in our first approximation (which we shall refer to as A1) we shall simply neglect the factors of $\gamma$ in the computation of both $\zeta$ and $\sigma_v$.

Neglecting such factors, the parameter $\zeta$ may now be estimated from Eq.~(\ref{zetadef}):
\be
\zeta = \frac{1}{\beta \tau_* f_1(s)}  = \frac{C(s)}{\beta} \sqrt{\frac{1-\lambda}{2\lambda+1}}   \,,
\label{eq:approxZeta}
\ee
where
\be
f_1(s) = \frac{\sqrt \pi \Gamma(\frac{s}{2}+1)}{2\Gamma(\frac{s+3}{2})}\,.
\ee
and
\be
C(1)=\frac{4}{\pi}\,, \qquad C(2)=\frac{3}{\sqrt{2}}\,,
\ee
for cylindrical and spherical domain walls, respectively. The value of $\beta$, corresponding to a fixed $\zeta$, is different for cylindrical and spherical domain walls. In fact
\be
\frac{\beta_{\rm spherical}}{\beta_{\rm cylindrical}}= \frac{C(2)}{C(1)}= \frac{3\pi \sqrt 2}{8} \sim 1.67
\ee

The root-mean-squared velocity of the domain walls, computed using Eqs.~(\ref{svdef}),~(\ref{analytic1a}),~(\ref{analytic1}) and~(\ref{analytic2}) and neglecting the factors of $\gamma$, is equal to
\be
\sigma_v^2 = \frac{1}{\tau_*^2} \frac{f_2(s)}{f_1(s)} =  \frac{1}{\tau_*^2}  \frac{\Gamma(\frac{s}{2})}{2\Gamma(\frac{s}{2}+1)} = \frac{1}{s\tau_*^2}=\frac{(1-\lambda)}{2\lambda+1} \,,
\label{eq:approxSv}
\ee
with
\be
f_2(s) =  \int_0^{1} x^2 (1-x^2)^{s/2-1} dx = \frac{\sqrt \pi\Gamma(\frac{s}{2})}{4 \Gamma(\frac{s+3}{2})} \,.
\ee

It is also possible to compute $\zeta$ and $\sigma_v$ using Eqs. (\ref{zetadef}) and (\ref{svdef}) considering a second approximation where the factors of $\gamma$ are no longer neglected [including the factor of $\gamma^{1/s}$ in Eq. (\ref{analytic1a})]. Although there is no simple analytical expressions for $\zeta$ and $\sigma_v$ in this case, the main difference is that the degeneracy between the results obtained considering cylindrical ($s=1$) and spherical ($s=2$) domain walls is broken. Since this is an important feature of the parameter-free VOS model, we shall come back to this approximation --- which we shall refer to as A2 --- in the following section.

\subsubsection{Curvature dominated evolution of individual domains (2D)}

For $\lambda$ sufficiently close to unity, the first term in the velocity evolution equation [given in Eq. (\ref{dwev})] may be neglected in the non-relativistic regime. Hence, for $s=1$ 
\be
v =  \frac{\eta}{q} \frac{1-\lambda}{3\lambda}\,, \label{vcyl}
\ee
to an excellent approximation, throughout almost the whole evolution of the domain walls. Although Eq. (\ref{vcyl}) describes the two-dimensional dynamics of the circular transversal cross section of a cylindrical domain wall, it may be easily generalized to account for a transversal cross section with an arbitrary shape:
\be
v =  \kappa \eta \frac{1-\lambda}{3\lambda}\,.
\ee
Here, $\kappa$ represents the corresponding comoving curvature at each point. The evolution of the cross-sectional transverse comoving area $\mathcal{A}$ of each domain wall with the conformal time $\eta$ is then given by 
\be
\dot {\mathcal A} = -\oint v dl = -  \frac{1-\lambda}{3\lambda} \eta \int \kappa dl = -  2\pi  \frac{1-\lambda}{3\lambda} \eta\,, \label{dotA}
\ee
independently of the shape of the cross-sectional transverse area. Here, $dl$ is a comoving infinitesimal arc length. Equation (\ref{dotA}) represents the particular case of von Neumman's law \cite{vonNeumann:1952:DRC,Avelino:2010qf} where the number of domain edges is equal to zero. The solution of  Eq. (\ref{dotA}) may be written as 
\be
\mathcal A = {\mathcal A}_{\rm i} - \pi  \frac{1-\lambda}{3\lambda} \eta^2\,,
\ee
where ${\mathcal A}_{\rm i}$ is the initial cross-sectional transverse comoving area $\mathcal A$ of the domain wall. This result implies that, for $\lambda$ sufficiently close to unity, the evolution of the area enclosed by a two-dimensional domain is essentially independent of its shape. 

\subsubsection{Curvature dominated evolution of individual domains (3D)}

If, as happens in the $s=2$ case, both the principal curvatures of the domain walls are nonzero, the evolution of the comoving volume $V$ enclosed by each the domain wall with the conformal time $\eta$ is given by \cite{2007Natur.446.1053M}
\be
\dot V = -\oint v dA = -  \eta \frac{1-\lambda}{3\lambda} \int \mathcal K dA =  -  2\pi \mathcal L\,,
\ee
where $dA$ and $\mathcal K$ are respectively an infinitesimal surface element  and the mean comoving curvature of the domain wall at each point, and $\mathcal L$ is the comoving mean width of the comoving volume enclosed by the domain wall. For a fixed initial comoving volume $V_{\rm i}$, $\mathcal L$ generally depends on the domain wall shape, which implies that the evolution of $V$ is in general not the same for domain walls with the same $V_{\rm i}$ but different shapes. 

\subsection{The $\lambda=0$ case}

In the $\lambda=0$ case, the scale factor $a$ is constant and non-vanishing --- we shall take $a=1$, so that the physical and conformal times coincide. 
Also, equation~(\ref{dwev}) now implies that $\tilde q^s \gamma=1$ or, equivalently, that 
\be
\tilde q = (1-v^2)^{1/(2s)}\,.
\ee

\subsubsection{Cylindrical domain walls}

For cylindrical domain walls ($s=1$) one has
\be
\tilde q = (1-v^2)^{1/2}\,.
\ee
Substituting this into Eq.~(\ref{zetadef}) one finds that
\be
\zeta=\frac{1}{\beta \tau_*}\,.
\label{taustar}
\ee
Therefore, the solution of Eq.~(\ref{dwev}) is given by
\be
\tilde q = \cos \tau \,, \qquad v= \sin \tau\,,
\label{solseq1}
\ee
where we have assumed that $\tau=0$ when $\tilde q =1$ and $v=0$.
The collapse then occurs for 
\be
\tau=\tau_*=\frac{\pi}{2} \sim 1.57\,,
\label{taustar1}
\ee
when $\tilde q=0$, $v=1$ and $\gamma=\infty$. It is then possible to show, using Eqs.~(\ref{zetadef}),~(\ref{svdef}),~(\ref{taustar}),~(\ref{solseq1}) and~(\ref{taustar1}), that
\be
\zeta=\frac{2}{\pi \beta} \,, \qquad \sigma_v=\frac{\sqrt 2}{2} \sim 0.71\,. 
\ee

\subsubsection{Spherical domain walls}

For spherical domain walls ($s=2$), the solution of Eq.~(\ref{dwev}) is given by
\be
\tilde q =(1-v^2)^{\frac{1}{4}}\,, \qquad v= \sin \left( 2 \, {\rm am} \left(\tau \, | \,2  \right)\right)\,,
\label{solseq2}
\ee
where ${\rm am} \left(x \, | \,m  \right)$ is the Jacobi amplitude function, and we have again assumed that $\tau=0$ when $\tilde q =1$ and $v=0$. In this case, the collapse occurs for 
\be
\tau=\tau_*=F\left(\,\frac{\pi}{4}\left|\frac{}{}2 \right.\right) \sim 1.31\,, 
\label{taustar2}
\ee
again when $\tilde q=0$, $v=1$ and $\gamma=\infty$. Here $F(x\,| \, m)$ is the elliptic integral of the first kind. Then, using Eqs.~(\ref{zetadef}),~(\ref{svdef}),~(\ref{solseq2}) and~(\ref{taustar2}), one can show that
\be
\zeta=\frac{1}{\beta F(\pi/4\,| \, 2)} \,, \qquad \sigma_v = \sqrt{\frac{2}{3}}  \sim 0.82\,. 
\ee

\begin{figure} 
	         \begin{minipage}{1.\linewidth}  
               \rotatebox{0}{\includegraphics[width=1\linewidth]{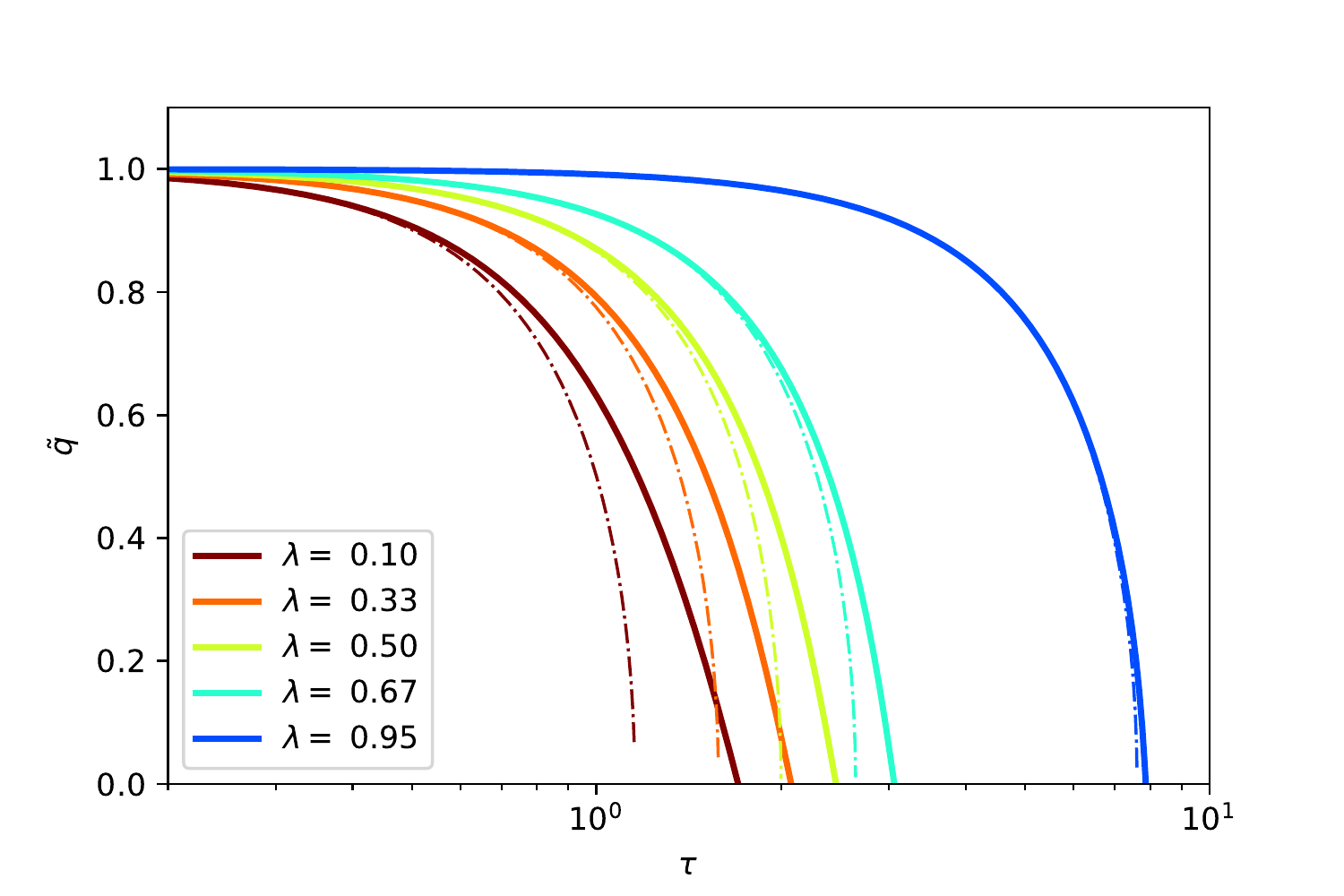}}
               \rotatebox{0}{\includegraphics[width=1\linewidth]{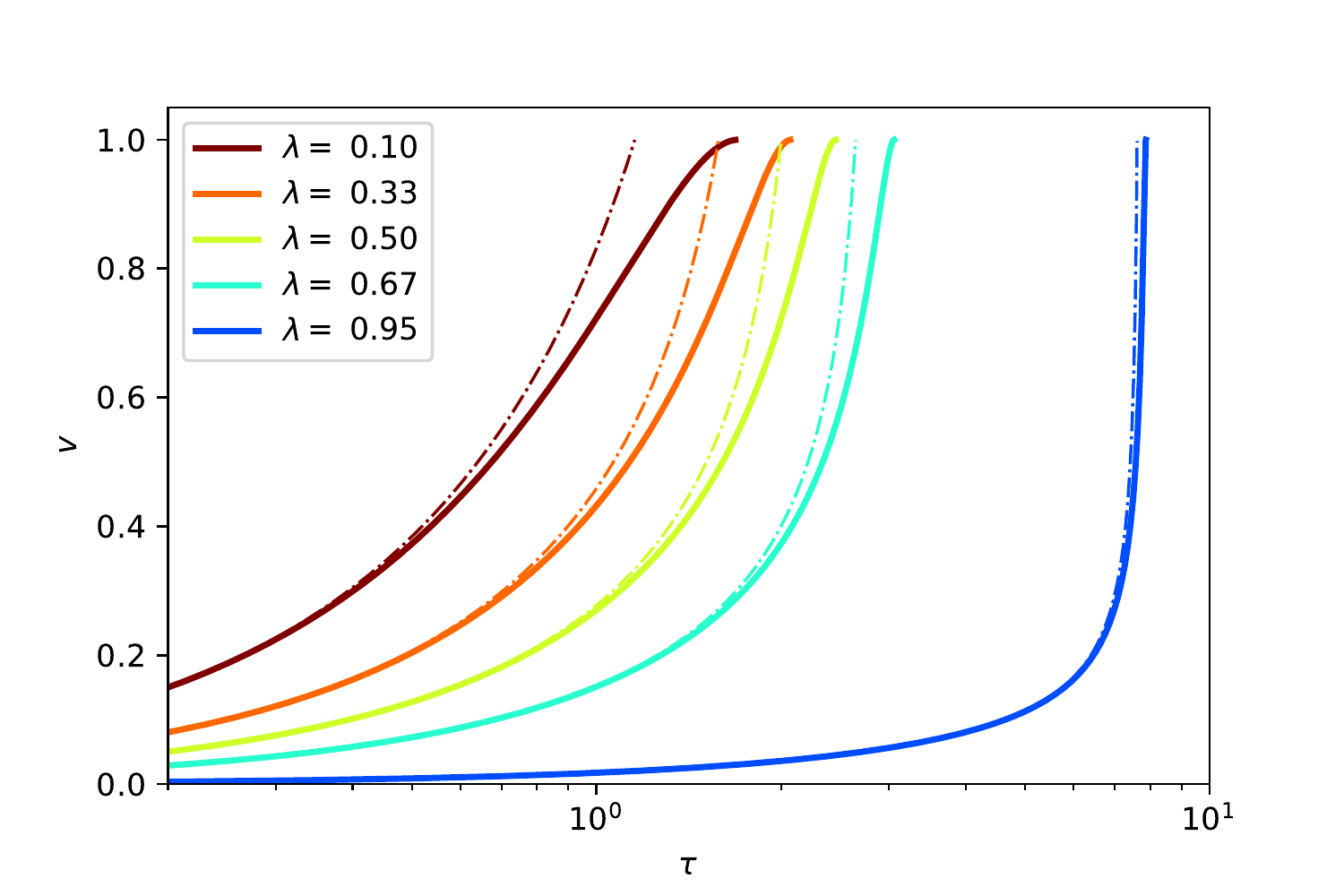}}
         \end{minipage}
	\caption{Collapse of cylindrical domain walls shown in terms of their reduced radius $\tilde{q}$ (top panel) and velocity $v$ (bottom panel) in universes with different expansion rates (parameterized by $\lambda$). The solid lines correspond to the numerical solution of the Nambu-Goto equations of motion while the dash-dotted lines represent the approximation given in Eqs. (\ref{analytic1a}), (\ref{analytic1}) and (\ref{analytic2}). Notice the excellent agreement between the analytical and numerical results, especially for $\lambda$ close to unity or low values of $\tau$.}
	\label{fig:qandvPlot}
\end{figure}

\section{Fitting the scaling predictions of the parameter-free VOS model}
\label{sec4}

In Fig. \ref{fig:qandvPlot} we show the evolution of the reduced radius $\tilde{q}$ (top panel) and velocity $v$ (bottom panel) of cylindrical domain walls in universes with different expansion rates, parameterized by $\lambda$. The solid line represents the numerical solution obtained using the Nambu-Goto equations of motion while the dash-dotted line represents the approximation given in Eqs. (\ref{analytic1a}), (\ref{analytic1}) and (\ref{analytic2}). Figure \ref{fig:qandvPlot} shows that, independently of the value of $\lambda$, the agreement between this approximation and the numerical solution is always excellent for sufficiently low values of $\tau$, in a regime where the domain walls are still non-relativistic. Also, for values of $\lambda$ close to unity, this excellent agreement spans almost the full conformal lifetime of the domain walls, since, as shown in the previous section, in this limit they only become relativistic extremely close to $\tau=\tau_*$. We have verified that qualitatively similar results are obtained in the case of spherical domain walls.

\begin{figure} 
	         \begin{minipage}{1.\linewidth}  

               \rotatebox{0}{\includegraphics[width=1\linewidth]{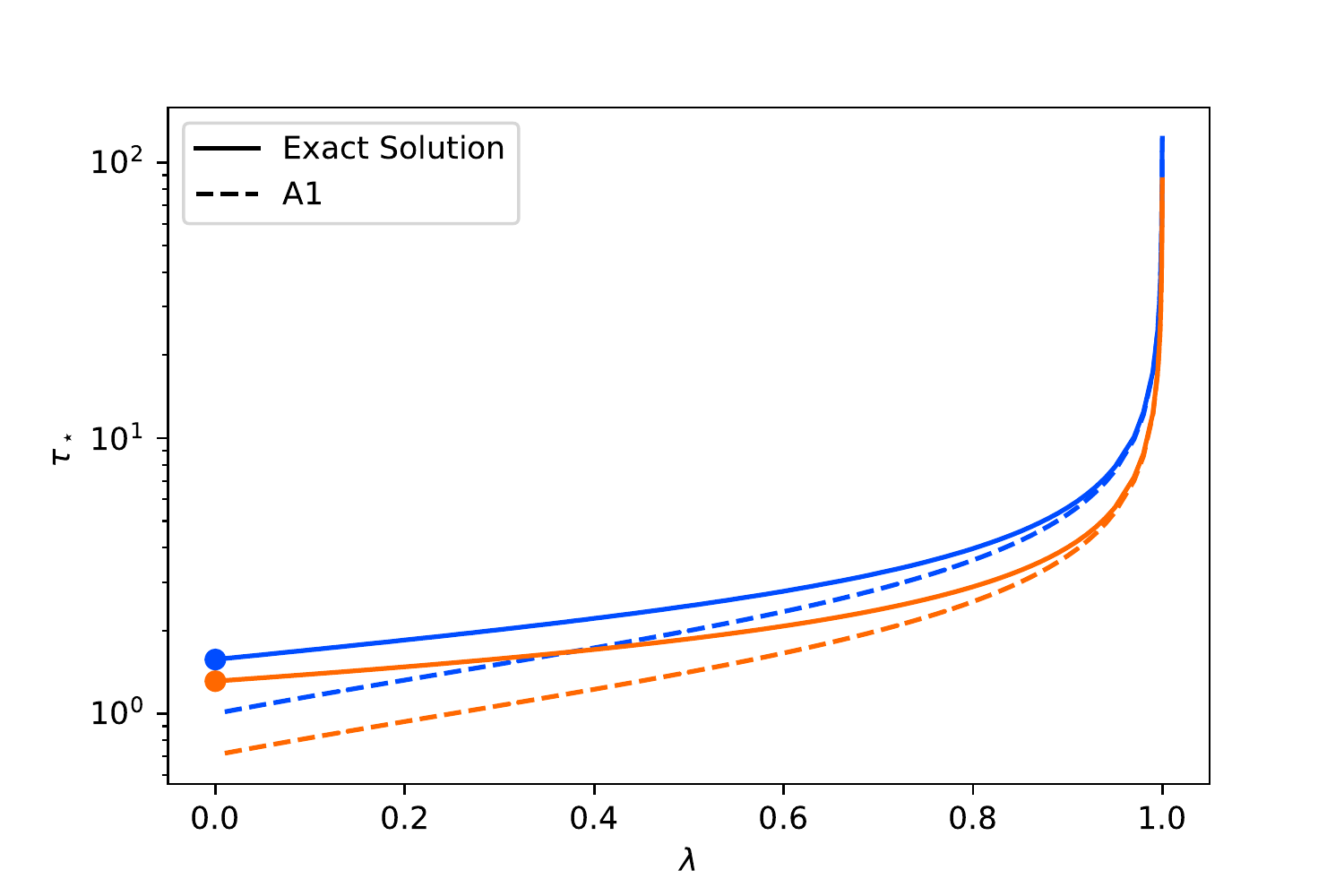}}
         \end{minipage}
	\caption{Comparison between the domain wall conformal lifetime, $\tau_*$, obtained by solving numerically the Nambu-Goto equations of motion (solid lines) and the analytic approximation given in \eqref{analytic2} (dashed lines). The blue and orange lines represent the results obtained considering cylindrical and spherical domain walls, respectively. Additionally, the exact solution in the $\lambda=0$ case is represented by the blue and orange circles. Notice the excellent agreement between the analytical and numerical results for $\lambda \gsim 0.9$.}
	\label{fig:taustar}
\end{figure}

In Fig. \ref{fig:taustar} we compare the value of the domain wall lifetime, parameterized by $\tau_*$, calculated by solving numerically the Nambu-Goto equations of motion (solid lines) with the prediction of the analytical approximation given in \eqref{analytic2} (dashed lines). The blue and orange lines display the results obtained for cylindrical and spherical domain walls, respectively, while the blue and orange circles represent the corresponding exact solutions for $\lambda=0$. Figure \ref{fig:taustar} shows that there is an excellent agreement between the analytical approximation and the  numerical solution for $\lambda \gsim 0.9$. On the other hand, for smaller values of $\lambda$ the approximation slightly underestimates the value of $\tau_*$.

\begin{figure} 
	         \begin{minipage}{1.\linewidth}  
               \rotatebox{0}{\includegraphics[width=1\linewidth]{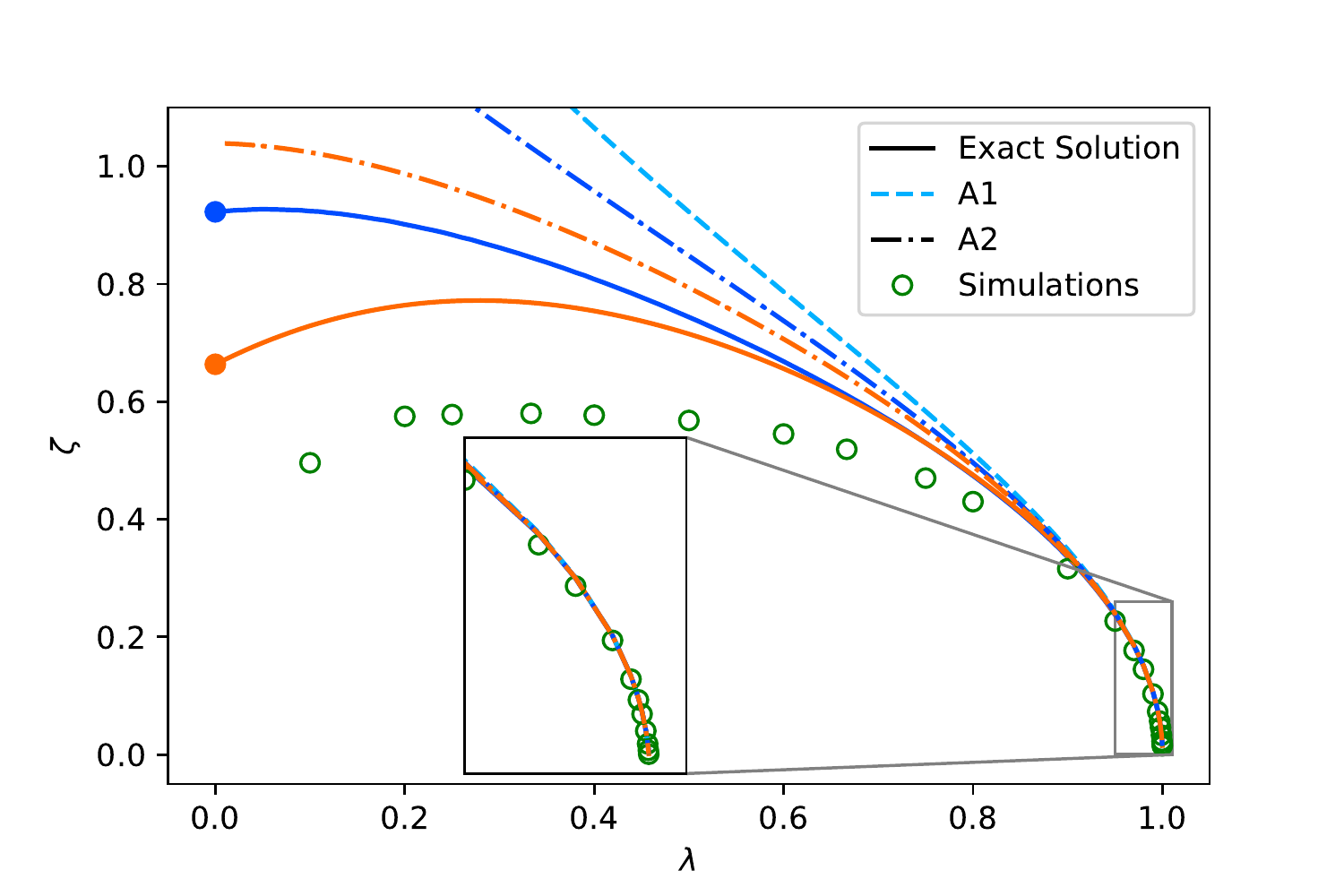}}
               \rotatebox{0}{\includegraphics[width=1\linewidth]{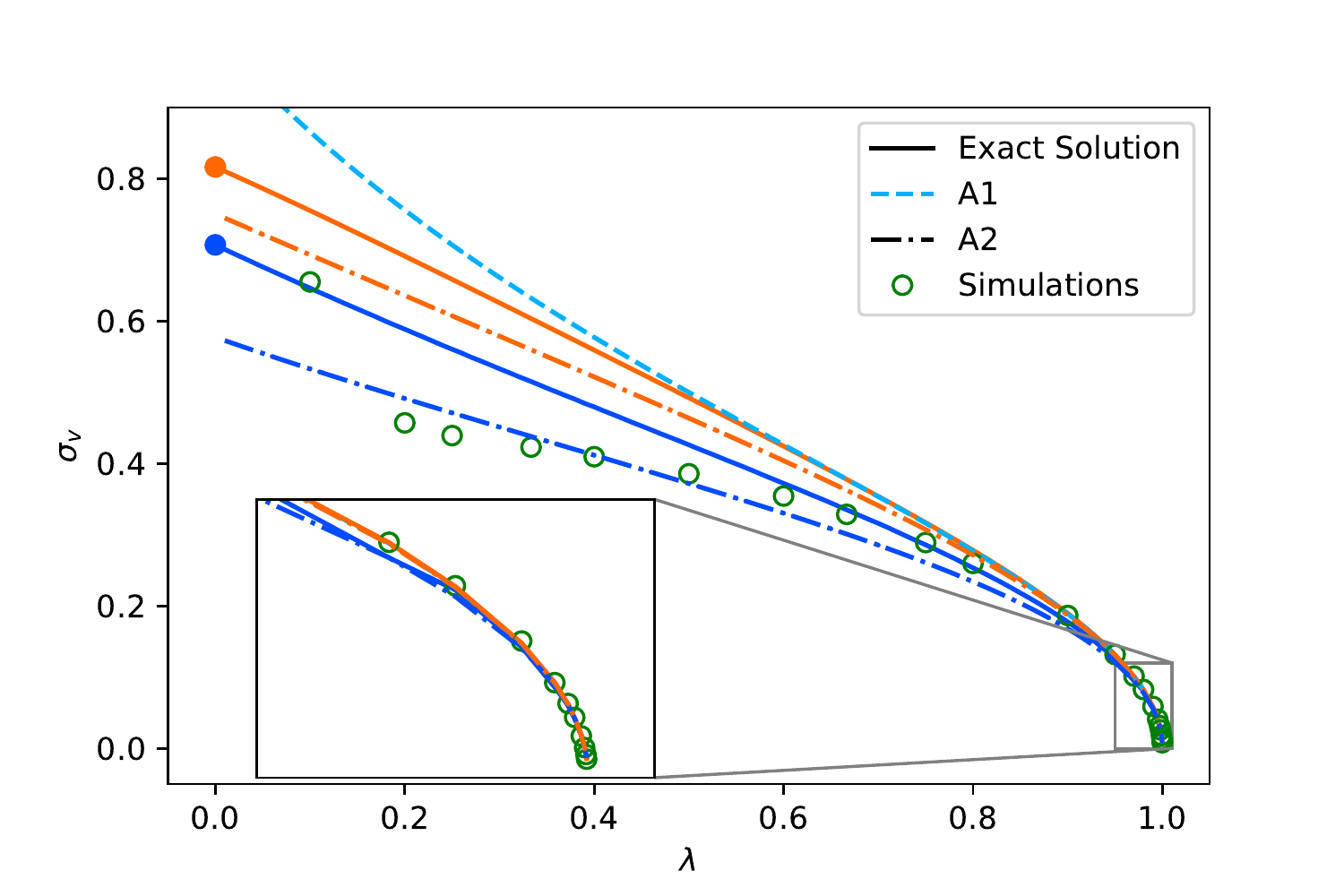}}
         \end{minipage}
	\caption{Summary of the different results obtained for $\zeta$ (top panel) and $\sigma_v$ (bottom panel). The solid lines represent the prediction of the parameter-free VOS model, which has an exact analytic solution for $\lambda=0$ (represented by the filled dots). The dashed and dash-dotted lines represent approximations A1 and A2, respectively. The unfilled dots show the results of field-theory simulations \cite{Martins:2016ois}. Notice the excellent agreement for $\lambda \gsim 0.9$ and that the geometrical degeneracy of the approximation A1 (blue dashed line) is broken in the approximation A2 (dashed blue and orange lines, respectively, for cylindrical and spherical domain walls).}
	\label{fig:zetaSigmav}
\end{figure}

Figure \ref{fig:zetaSigmav} summarizes the different results obtained for $\zeta$ (top panel) and $\sigma_v$ (bottom panel). The solid lines show the (numerically evaluated) prediction of the parameter-free VOS model for cylindrical and spherical domain walls (in blue and orange, respectively), which has an exact analytic solution for $\lambda=0$ (represented by the blue and orange circles). The dashed and dash-dotted lines represent, respectively, the approximations A1 and A2 discussed in the previous section. Here, we have taken $C/\beta = 1.85$, thus requiring the model to reproduce the value of $\zeta$ obtained using field theory simulations for $\lambda=0.9998$ \cite{Rybak2018}.

Figure \ref{fig:zetaSigmav} shows that both approximations do an excellent job for $\lambda \gsim 0.8$, but deviate more significantly from the model prediction for smaller values of $\lambda$. Also notice that, while approximation A1 predicts the same values of $\zeta$ and $\sigma_v$ for both cylindrical and spherical domain wall configurations, the degeneracy between these two different geometries is broken by approximation A2. This is indeed a property shared by the exact solutions, thus showing that the geometry of the domain walls plays a crucial role for small values of $\lambda$ in which the relativistic regime of the domain wall evolution lasts for longer. 

The unfilled circles in Fig. \ref{fig:zetaSigmav} represent the values of $\zeta$ and $\sigma_v$ measured in field-theory numerical simulations \cite{Martins:2016ois}. Notice the excellent agreement for $\lambda \gsim 0.9$ between the predictions of the parameter-free VOS model and the simulation results (see also Table \ref{table1}). For smaller values of $\lambda$ there is still good qualitative agreement, but significant discrepancies occur. These discrepancies may be associated with the limitations of numerical field theory simulations in the relativistic limit~\cite{Hindmarsh:2014rka} and of the methods used to extract $L$ and $\sigma_v$ or may stem from the fact that the parameter-free VOS model only considers simple geometries. These are discussed in detail in \cite{Avelino:2019wqd,Avelino:2020ubr} and will need to be addressed in future work. An important feature that a comparison between the prediction of the parameter-free VOS model and the results of numerical simulations will have to explain is the fact that both the results for $\zeta$ and $\sigma_v$ fall below the model predictions.

The analytical results obtained in the present paper can be used to construct an accurate fit to the scaling predictions of the parameter-free VOS model for $\zeta$ and $\sigma_v$ over the whole range of $\lambda$ ($\lambda \in [0,1[$) for which a linear scaling solution with cosmic time is possible in an expanding homogeneous and isotropic FLRW universe. We shall consider a fit of the form
\begin{equation}
\mathcal F (\lambda,s) = (1-\lambda)^n  \mathcal F_{\lambda=0}(s) + \lambda^m \mathcal F_{\rm A1}(s) \, ,
\end{equation}
where $\mathcal F$ represents either $\zeta$ or $\sigma_v$, $\mathcal{F}_{\lambda=0}(s)$ is the exact solution for $\mathcal F$ in the $\lambda=0$ case, $\mathcal{F}_{\rm A1}(s)$ is the first analytical approximation $A1$ for $\mathcal F$ as given in Eqs. \eqref{eq:approxZeta} and  \eqref{eq:approxSv}, $s=1,2$, and $n, \, m$ are the fit parameters given in Table \ref{table2}. We have verified that the maximum relative error of the fits for both $\zeta$ and $\sigma_v$ is of the order of $1 \%$, allowing us to provide a closed formula to the scaling values of $\zeta$ and $\sigma_v$ predicted by the parameter-free VOS model.

\begin{table}[t]
\begin{tabular}{|c| c c |  c c |}
\hline
$\lambda$ & \multicolumn{2}{|c|}{$\zeta$} & \multicolumn{2}{|c|}{$\sigma_v$} \\ 

& A1  & Simulations & A1 &  Simulations \\
\hline
0.97 & 0.186  & 0.177 & 0.101  & 0.102 \\
0.98 & 0.152  & 0.145 & 0.082  & 0.083 \\
0.99 & 0.107  & 0.103 & 0.058  & 0.059 \\
0.995 & 0.075  & 0.073 & 0.041  & 0.041 \\
0.997 & 0.058  & 0.057 & 0.032  & 0.032 \\
0.998 & 0.048  & 0.046 & 0.026  & 0.026 \\
0.999 & 0.034  & 0.033 & 0.018  & 0.018 \\
0.9995 & 0.024  & 0.023 & 0.013  & 0.013 \\
0.9997 & 0.018  & 0.018 & 0.010  & 0.010 \\
0.9998 & 0.015  & 0.015 & 0.008  & 0.008 \\
\hline
\end{tabular}
\caption{The values of $\zeta$ and $\sigma_v$ obtained using approximation A1 (which follow very closely the results of the parameter-free VOS model for $s=1$ and $s=2$) and the results of field theory numerical simulations for discrete values of $\lambda$ between $0.97$ and $0.9998$. Notice the exceptional agreement between approximation A1 (and consequently the predictions of the parameter-free VOS model) and the results of field theory numerical simulations in the $\lambda \to 1^-$ limit.}
\label{table1}
\end{table}

\begin{table}[t]
\begin{center}
\begin{tabular}{c | c c }
($n, \, m$) & \text{cylindrical} & \text{spherical} \\
\hline
$\zeta$ & ($1.6371, \, 1.0413$) & ($1.4044, \, 0.9975$) \\
$\sigma_v$ & ($1.7947, \, 1.1686$) & ($1.6179, \, 1.1346$) \\

\end{tabular}
\end{center}
\caption{Parameters $(n,m)$ of the fits for $\zeta$ and $\sigma_v$ for $s=1$ (cylindrical domain walls) and $s=2$ (spherical domain walls).}
\label{table2}
\end{table}

\section{Parametric VOS model}
\label{sec5}

In this section, we will consider the non-relativistic limit of the original (parametric) VOS model for the cosmological evolution of domain wall networks.
This model provides a quantitative thermodynamical description of their cosmological evolution by following the evolution of $\sigma_v$ and $L$~\cite{Avelino:2010qf,Avelino:2011ev,Sousa:2011iu}:
\begin{eqnarray}
\frac{d\sigma_v}{dt} & = & \left(1-\sigma_v^2\right)\left[\frac{k}{L}-\frac{\sigma_v}{\ell_d}\right]\,,\label{VosS}\\
\frac{dL}{dt} & = & HL+\frac{L}{\ell_d}\sigma_v^2+{\tilde c}\sigma_v\,.\label{VosL}
\end{eqnarray}
Here, we introduced the damping length scale and $\ell_d^{-1}=3H+\ell_f^{-1}$, which incorporates both the impact of the cosmological expansion (through the Hubble parameter $H$) and the effect of the frictional forces caused by the scattering of particles off the domain walls (encoded in the friction length scale, $\ell_f$). In what follows, we shall neglect the effects of friction in the evolution of the network (i.e., we set $\ell_f=+\infty$).

Notice that this model has two parameters that need to be calibrated against numerical simulations. The first of these, $k$, is a dimensionless curvature parameter that, to some extent, describes the average domain wall curvature. The second parameter, $\tilde c$, was introduced to quantify the efficiency of the energy loss associated with the collapse of the domain walls. This energy loss results in a contribution to the evolution of the average domain wall energy density which, analogously to the cosmic strings' case \cite{Kibble:1984hp}, is usually assumed to be of the form
\be
\left.\frac{d\rho_{\rm w}}{dt}\right|_{+}= \tilde c\, \sigma_v \frac{ \rho_{\rm w}}{L}\,.
\label{energyloss}
\ee
For infinitely-thin and featureless domain walls, Eqs.~(\ref{VosS}) and (\ref{VosL}) may be obtained directly from the generalized Nambu-Goto action assuming a FLRW background and making a couple of approximations, except for the $\tilde c\, \sigma_v / D$ term that results from this energy loss~\cite{Sousa:2011iu}. 

Conservatively, we may expect the rate of variation of energy not associated to the expansion of the cosmological background to be such that
\be
\left.\frac{d\rho_{\rm w}}{dt}\right|_{+}= \tilde c \sigma_v\frac{\rho_{\rm w}}{L} \lsim  \frac{\rho_{\rm w}}{\Delta t} \label{drhodtp2}\,,
\ee
where $\Delta t$ is the physical time required for a domain wall to travel across a comoving distance of $L/a$. We should then have that
\be 
{\tilde c}\lsim \frac{L}{\sigma_v\Delta t}\,.
\label{tildec}
\ee 

Domain wall networks are known to evolve towards a linear scaling regime of the form
\be
L=\xi t \,, \qquad \sigma_v= \rm const\,,
\ee
for a power law expansion of the form $a \propto t^\lambda$, with constant $\lambda$ and $0 <  \lambda < 1$. The parametric VOS model predicts that this regime should be characterized by: 
\be
\xi=\sqrt{\frac{k(k+\tilde c)}{3\lambda(1-\lambda)}}\quad\mbox{and}\quad \sigma_v=\sqrt{\frac{(1-\lambda)k}{3\lambda(k+\tilde c)}} \label{linear}\,,
\ee
and thus these equations may be used to calibrate the free parameters $k$ and $\tilde c$ using numerical simulation. 

In the non-relativistic regime, with $\lambda\to 1^-$, this model then predicts that $\xi \to \infty$ and $\sigma_v \to 0$. This necessarily means that the energy loss term in Eq.~(\ref{energyloss}) is, given its dependence on $L$ and $\sigma_v$, irrelevant in this limit. In what follows, we will further show that we necessarily have that ${\tilde c}\to 0$ in the non-relativistic limit and that, as a result, the importance of this energy loss term decays significantly faster than Eq.~(\ref{energyloss}) seems to indicate as $\lambda\to 1^-$. Notice that
\be
\zeta=\frac{L}{a\eta}=(1-\lambda) \xi=\sqrt{\frac{k(k+\tilde c)(1-\lambda)}{3\lambda}} \to 0
\ee
in the same limit (here we have taken into account that $\eta=t/[(1-\lambda)a]$). 

The time $\Delta t$ a domain wall takes to transverse a comoving distance of $L/a$ may be estimated as
\be
\sigma_v \int_t^{t+\Delta t} \frac{dt} a = \frac{L}{a}=  \frac{\xi t}{a}\,,
\ee
and hence it is given approximately by
\be
\Delta t =\left( \left[ \frac{\xi}{\sigma_v}(1-\lambda)+1\right]^{1/(1-\lambda)}-1\right) t\,. \label{dt}
\ee
For $\lambda=0$, Eq. (\ref{dt}) yields
\be
\Delta t_{\lambda=0} = \frac{\xi t}{\sigma_v} = \frac{L}{\sigma_v} \,,
\ee
and, thus, Eq.~(\ref{tildec}) implies that we should have $\tilde c \le 1$. This shows that, in the relativistic limit, Eqs.~(\ref{tildec}) and ~(\ref{dt}) still leave some freedom as to the choice of $\tilde c$. However, in the non-relativistic limit this is not the case. Taking into account that Eq.~(\ref{linear}) implies that
\be
\frac{\xi}{\sigma_v}(1-\lambda) = \left|k+\tilde c\right|> 0\,,
\ee
one finds that, in the $\lambda\to 1^-$ limit, $\Delta t$ should be infinitely larger than $L/\sigma_v$. It then follows from Eq.~(\ref{tildec}) that $\tilde c$ must necessarily vanish in this limit. 

This actually means that, in the non-relativistic limit, the parametric VOS model must reduce to the parameter-free model as it effectively includes no additional source of energy loss (beyond expansion). In fact, in this limit, we should have
\be 
\sigma_v=\sqrt{\frac{1-\lambda}{3\lambda}}\quad\mbox{and}\quad\zeta=k\sqrt{\frac{1-\lambda}{3\lambda}}\,,
\label{vosnonrel}
\ee 
which has the same asymptotic behaviour as the approximation derived in Sec.~\ref{non-rel-app} using the parameter-free model \footnote{The approximation in Sec.~\ref{non-rel-app} was derived by taking into account the dominant term in the domain wall acceleration. A simpler approximation can be derived by neglecting the acceleration of the walls in this limit. In this case, domain wall evolution would still be described by Eqs.~(\ref{analytic1a})-(\ref{analytic1}), but with $\tau_*^2=3\lambda/(1-\lambda)$. This leads to an approximation similar to Eqs.~(\ref{eq:approxZeta}) and~(\ref{eq:approxSv}), but with the substitution $2\lambda+1 \to 3\lambda$. 
}. 
Notice that $k$ in Eq.~(\ref{vosnonrel}) and $C/\beta$ in Eq.~(\ref{eq:approxZeta}) are equivalent in the $\lambda \to 1^-$ limit.

This coincidence between both models in the $\lambda\to 1^-$ limit is in agreement with the conclusions of~\cite{Avelino:2020ubr}, where the parametric and parameter-free models were compared in detail. Therein they found that the parametric model overestimates the strength of the Hubble damping term --- as it does not account for the dispersion of $\sigma_v$ --- and that the impact of wall decay on $\sigma_v$ is not included. In the non-relativistic limit, however, these two effects are expected to be negligible and thus the two models should be equivalent in the $\lambda\to 1^-$. Furthermore, the results of~\cite{Avelino:2020ubr} indicate that 
$\tilde c\to 0$ in the $\lambda \to 1^-$ limit, in agreement with our previous discussion. The dependence of $\tilde c$ on the rate of expansion seems to suggests that the scaling of the energy loss term with $\sigma_v$ and $L$ (which depend on $\lambda$ themselves) should be different from that given in Eq.~(\ref{energyloss}).

The fact that we necessarily have that $\tilde c \to 0$ in the $\lambda\to 1^-$ limit further implies that, if one uses different numerical simulations with a wide range of expansion rates to calibrate the parameters of the original VOS model, one will necessarily find that no unique calibration exists. This is precisely what was found in~\cite{Martins:2016ois}: although in the relativistic limit simulations are well described by a $\tilde c$ close to unity, for fast expansion rates simulations are compatible with $\tilde{c}=0$. Given these results, the authors proposed an extension of the VOS model, involving four additional free parameters, that provides a good fit to the results of their simulations. This multi-parameter VOS model drops the assumption that $k$ is a constant --- assuming that $k(\sigma_v)$ instead --- and includes another energy loss term --- a ``radiation'' term that depends on $k(\sigma_v)$ ---  while keeping the term in Eq.~(\ref{energyloss}) with a constant $\tilde c$. This multi-parameter model, however, reduces to the original VOS model in the non-relativistic limit (since the additional energy loss term becomes irrelevant in this limit and $k$ becomes approximately constant) and, as a result, calibration with simulations, unsurprisingly, yielded the $\tilde c=0$.

\section{Conclusions}
\label{sec6}

In this paper we investigated the dynamics of standard frictionless domain wall networks in Friedmann-Lemaître-Robertson-Walker universes considering a recently proposed parameter-free VOS model. We derived an analytical approximation for the scaling evolution of the two thermodynamic parameters of the model, $L$ and $\sigma_v$, for models with $\lambda$ close to unity, showing that it becomes exact in the $\lambda \to 1^-$ limit. We have also computed an exact solution valid in the $\lambda=0$ limit. We used these results to construct a fit of the model predictions valid for $\lambda \in [0, 1[$ with a maximum error of the order of $1 \%$. We have shown that this fit is not only in good qualitative agreement with the results of field theory simulations over the whole range of $\lambda$'s considered in the present paper, but also that the quantitative agreement is exceptional for values of $\lambda$ close to unity despite the extremely fast variation of the two parameters in this limit. We have further demonstrated that the phenomenological energy loss parameter of the standard VOS model and of a recently proposed multi-parameter one-scale domain wall model vanish in the $\lambda \to 1^-$ limit.

The analytical tools developed in the present paper may prove to be very useful in future studies of the cosmological consequences of domain wall networks, since they may enable the development of analytical frameworks to characterize various observational signatures. The derivation of observational constraints, in particular, often requires multiple computations of their observational signatures over a wide parameter space, which may not only be computationally costly but also time-consuming when based on numerical studies. These results, however, may provide a quick and versatile alternative to derive observational constraints on different domain wall scenarios using the data of multiple probes.

\begin{acknowledgments}
L. S. is supported by FCT - Funda\c{c}\~{a}o para a Ci\^{e}ncia e a Tecnologia through contract No. DL 57/2016/CP1364/CT0001. D. G. is supported by FCT through the PhD fellowship 2020.07632.BD. Funding for this work has also been provided by FCT through national funds (PTDC/FIS-PAR/31938/2017) and by FEDER—Fundo Europeu de Desenvolvimento Regional through COMPETE2020 - Programme for Competitiveness and Internationalisation (POCI-01-0145-FEDER-031938), and through the research grants UIDB/04434/2020 and UIDP/04434/2020.

\end{acknowledgments}
 
\bibliography{walls}
 	
 \end{document}